# Transitioning from Blackboard to Moodle amidst Natural Disaster: Faculty and Students Perceptions

Ajayi Ekuase-Anwansedo
Southern University and A & M
ajayi_anwansedo_00@subr.edu

Jose Noguera
Southern University and A & M
jose_noguera@subr.edu

Brandon Dumas
Southern University and A & M
brandon_dumas@subr.edu

## ABSTRACT

Higher educational institutions continuously look for ways to improve the quality of their eLearning services and adapt learning solutions to suit the needs of the institution. During the 2016 Fall Semester, a university located in the Southern part of United States decided to transition from the Blackboard learning management system (LMS) to the Moodle learning management system. Typically such a transition presents a huge challenge for the University staff, faculty, and students. Additionally, on August 2016, what CNN themed "the worst natural disaster, to strike the United States since Hurricane Sandy" [47], occurred in Louisiana during the transition. This led to massive disruptions in activities throughout the state.

This paper examines the perceptions of both faculty and student on the transition from one LMS to another and also what impact, if any, the natural disaster had on the process. Faculty and students were surveyed to gain understanding of how they perceived the transitioning process, their perception of both systems, their preferences, and why. Furthermore, we identified issues peculiar to transitioning during a natural disaster. The results of this study can be used to anticipate issues that may be associated with transitioning from one LMS to the other and issues peculiar to transitioning amidst a natural disaster. It can also be used to identify areas for improvement.

## 1 INTRODUCTION

As information technology is being integrated into teaching and learning in our higher educational institutions [20], learning management systems (LMSs) have become the norm. This has led to the development of a variety of LMSs, both commercial and non-commercial. The non-commercial LMS are called open source and most of them allow the user to obtain the software and adapt it to suit their needs. Previous research indicated that Moodle is the most popular open source LMS [5], [21] and is preferred over Blackboard by students [8], [9], [29]. Some of the advantages of Moodle include adaptability, its design philosophy based on social constructionist pedagogy [11], absence of license and maintenance fees [11], [43], strong community of users [11], [46] and providing self-directed learning to students [26]. Even though, Moodle has rave reviews, its success is dependent on its acceptance and use by the students. [3], [24], [25], [36], [48]. Hence, there is a need to investigate the adoption of Moodle by students especially in unusual circumstances, to determine its success and if these external factors influence its adoption. The technology acceptance model (TAM) proposed by Davis, [14], has been identified by Hsu & Chang, as the most used tool for predicting technology acceptance and usage intentions [24]. In this study, a research model uses TAM as a base model and integrates the impacts of a natural disaster, in this instance, a flood, as external variables, to examine how it affects faculty and students' acceptance behavior and use intentions of Moodle.

The remainder of this paper will first review research on TAM and TAM extensions, the impact of flood, the impact of emotion on cognition in PTSD, Depression and Anxiety. An extended model of TAM that presents the impact of the flood on Moodle adoption is then developed. Finally contributions of this study and suggestions for future research are discussed.

## 2 PURPOSE OF STUDY

Transitioning from one LMS to another presents unique challenges of integrating and adapting to the new technology. However, when this transition takes place during or after a natural disaster, it presents a whole new challenge that includes not only accepting and using the new technology but also managing the after effects of the natural disaster.

Amidst the transition process from Blackboard to Moodle learning management system by a university in the southern part of the United States, one of the worst natural disasters in the United States occurred [47]. Thus in addition to investigating the adoption of Moodle among the faculty and students, there is a need to examine the impact of the flood on faculty and students' acceptance and use of Moodle.

Flooding impacts individuals economically, medically, socially, and mentally; which in turn impacts cognitive function and consequently behavior. In addition, flooding has been the most common type of natural disaster globally and there is a likelihood of increased flooding in the future due to rising sea levels and more frequent and extreme precipitation [1], [2]. Also, during natural disasters, where physical commute is impossible, virtual learning presents an opportunity for continued, uninterrupted teaching and learning activities [25]. Consequently, there is a need to investigate how natural disasters influences the adoption and use of technology in this case Moodle.

A similar study [25] investigated factors which may affect the intention to use Moodle by university students at a university in the Eastern Region of Turkey after an earthquake. They extended TAM to include technical support and computer self-efficacy. However, they did not examine if the psychological impact of the earthquake on the university students had any effect on their intention to adopt or use Moodle.

Thus, this study examines the psychological impact of natural disasters - flood on technology adoption and use, in this instance Moodle.

## 3 RESERCH MODEL AND HYPOTHESIS FOR MOODLE ADOPTION AMIDST A NATURAL DISASTER - FLOOD

In this study, the research model proposed is based on TAM with a view to investigate the acceptance and use of the Moodle by faculty and students in, the university.

In our model, we extended TAM to include external factors – PTSD, depression, and anxiety -- which are the most common impact of a flood -- to examine faculty and student adoption of Moodle in the midst of a flood event at the University. The research model, as illustrated in Figure 2, consists of six constructs: PTSD, depression, and anxiety; perceived usefulness (PU); perceived ease of use (PEU); attitude towards use (ATU); behavioral intention to use (BIU); and Actual system use (ASU).

### 3.1 Technology Adoption Model (TAM)

The technology acceptance model (TAM), developed by Davis [14] is the most researched model used to determine the perception of user's acceptance and use of technology [25], [40]. It is based on the premise that the success of technology is based on the user's acceptance of that technology which is dependent on the perceived use, the perceived ease of use and the intention to use the technology [3], [15]. According to TAM, an individual's perception of the effort involved in using the technology and how much they think the technology improves their work will help them determine if they want to use the system or not [15]. In other words, the decision to use a technology is largely dependent on the individual's behavioral intention to use the technology, which is formed as a result of a cognitive process [37], [45].

The TAM is made up of six constructs (figure 1) namely: perceived ease of use (PEOU), perceived use (PU), attitude towards use (ATU), behavioral intention to use (BIU), actual system use (ASU) and external variables (EV). These six constructs help determine user's perception towards adopting a new technology. In the context of Moodle as the technology, PEOU represents the individual's belief that Moodle will require minimum effort to use, PU represents the individual's belief that using Moodle will enhance their work [15], [34]. ATU represents the individual's desire to use or not to Moodle, BIU represents the user's willingness to use Moodle, ASU represents the use of Moodle by the individual and EV comprises system design, user characteristics, environments and user involvement [10], [34], [37], [45].

TAM proposes that PEOU, PU and ATU determines individuals' attitude towards using a system, PEOU and PU has a significant impact on BIU, which influences BIU, which invariably determines the actual usage of the system [15], [21], [34], [40], [48].

Therefore, the following hypotheses based on TAM are proposed:

[H1] Perceived ease of use (PEOU) has a positive effect on the perceived usefulness (PU) of Moodle.

[H2] Perceived ease of use (PEOU) has a positive effect on attitudes toward the use (ATU) of Moodle.

[H3] Perceived ease of use (PEOU) has a positive effect on the behavioral intention to use (BIU) Moodle.

[H4] Perceived usefulness (PU) has a positive effect on attitudes toward the use (ATU) of Moodle.

[H5] Perceived usefulness (PU) has a positive effect on the behavioral intention to use (BIU) Moodle.

[H6] Attitudes toward using (ATU) have a positive effect on the behavioral intention to use (BIU) Moodle.

[H7] Attitude towards use (ATU) has a positive effect on the actual use (ASU) of Moodle.

[H8] Behavioral intention to use (BIU) has a positive effect on the actual use (ASU) of Moodle.

### 3.2 External variables

External variables are essential in strengthening the TAM model [24].

Several researchers have suggested that external variables influences the user acceptance of a system via the cognitive constructs (PEOU and PU) of the TAM model [3], [10], [14], [15], [45].

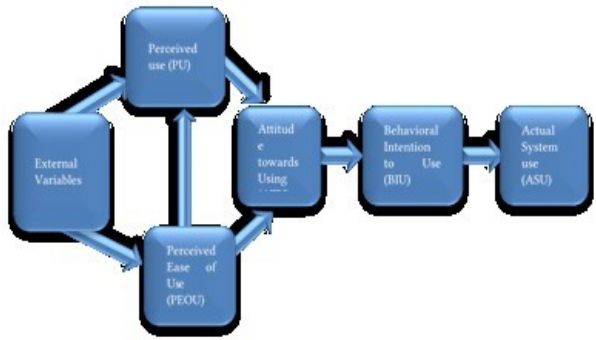

Figure 1: Technology Acceptance Model (Adapted from Davis, 1986)

## 3.3 Impact of the flood

Natural disasters are unintended traumatic events [27], [32]. Globally, floods are the most common natural disaster and is accountable for almost half of all natural disaster fatalities [2], [16], [19]. In addition to economic losses, other harmful impacts of flooding include mortality, mental health issues, damage to physical health, infrastructure and property [16], [19], [22]. Furthermore, post-traumatic stress disorder (PTSD), depression and anxiety, have been identified as the most commonly reported psychological effects of flooding [7], [19], [22], [31]. When compared to non- flooded areas higher levels of these psychological effects have been observed in flood affected areas [17], [41].

Also, exposure to floods have negative impacts on cognition [30]. Given that flood events are likely to increase in frequency and intensity in the future, due to climate change and exposure to flood events is likely to increase due to population growth, population proximity to coastline and increasing urbanization, there is a need to increase flood prevention and mitigation strategies [2], [16] in order to avoid been casualties of potential flood events [33].

## 3.4 Impact of emotions on cognition in Posttraumatic stress disorder (PTSD), Depression and Anxiety

PTSD is a mental health disorder which occurs after exposure to traumatic or stressful events, Depression is described as a feeling of intense sadness, hopelessness and worthlessness and Anxiety is the anticipation of future treat [4]. PTSD, Depression and Anxiety has been associated with some form of cognitive dysfunction such as impaired concentration, impaired decision-making ability and memory impairment [4], [23], [13], [44] and has significant impact on cognitive structure –the way the individual stores information internally, cognitive operations – the individual's information processing processes and cognitive products – what information the individual recalls [1]. In other words, individuals with symptoms of PTSD, Depression or anxiety experience bias when processing information [1]. For instance, when given a task such individuals, focus more on task unrelated information (e.g. fear, worry) and this reserves a small space in the working memory for task related information [18]. In the context of adopting a new technology, in this instance, Moodle, PTSD, depression and anxiety will impact negatively the cognitive processes associated with using the technology and in the TAM model, this is represented by the PEOU and the PU [37], [42].

Therefore, we hypothesis that;

[9]    PTSD will have a negative effect on PEOU

[10]   PTSD will have a negative effect on PU

[11]   Depression will have a negative effect on PEOU   [12] Depression will have a negative effect on PU [13] Anxiety will have a negative effect on PEOU [14] Anxiety will have a negative effect on PU

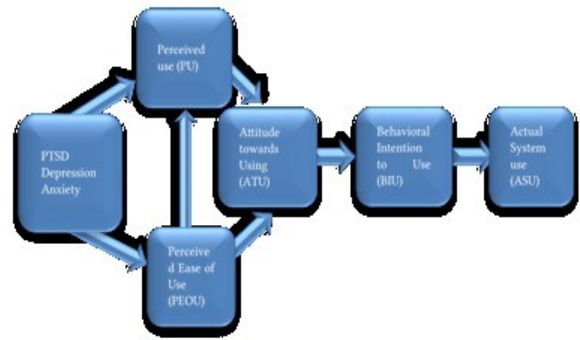

Figure 2: Extended TAM to include PTSD, Depression and Anxiety

## 4 FUTURE RESEARCH

In future research data will be collected by means of the following constructs, TAM (Davis, 1989), PTSD Checklist Civilian Version (PCLC) [12], – Center for epidemiologic studies depression scale (CES-D scale) [35] and the Generalized Anxiety Disorder 7 –item (GAD - 7) scale [38], [28] and analyzed using structural equation modeling (SEM)

## 5 CONCLUSION

Natural disasters are inevitable with the present climate conditions in the world today. Technology presents a way to mitigate the impact of natural disasters in education, by ensuring provision of uninterrupted educational services via learning management systems. This paper examines how a natural disaster may impact the adoption and use of technology. The conceptual model and 14 hypotheses was proposed integrating PTSD, depression and anxiety as external variable into the TAM model.